\documentclass[12pt,twoside]{article}
{\catcode `\@=11 \global\let\AddToReset=\@addtoreset}
\AddToReset{equation}{section}

\def\greaterthansquiggle{\raise.3ex\hbox{$>$\kern-.75em\lower1ex\hbox{$\sim$}}}
\def\lessthansquiggle{\raise.3ex\hbox{$<$\kern-.75em\lower1ex\hbox{$\sim$}}}

\newcommand{\beq}{\begin{equation}}
\newcommand{\eeq}{\end{equation}}
\newcommand{\beqa}{\begin{eqnarray}}
\newcommand{\eeqa}{\end{eqnarray}}
\newcommand{\beqan}{\begin{eqnarray*}}
\newcommand{\eeqan}{\end{eqnarray*}}
\newcommand{\ba}{\begin{array}}
\newcommand{\ea}{\end{array}}
\newcommand{\no}{\nonumber}

\newcommand{\sgn}{\rm sign}
\newcommand{\Det}{\rm Det\,}
\newcommand{\sh}{\rm sh}

\newcommand{\Un}{\underline}

\newcommand{\ra}{\rightarrow}

\newcommand{\ve}{\varepsilon}
\newcommand{\vp}{\varphi}

\newcommand{\wt}{\widetilde}

\newcommand{\A}{{\cal A}}

\newcommand{\C}{{\cal C}}

\newcommand{\dsum}{\displaystyle \sum}
\newcommand{\dprod}{\displaystyle \prod}

\def\nz{\ifmmode {I\hskip -3pt N} \else {\hbox {$I\hskip -3pt N$}}\fi}
\def\zz{\ifmmode {Z\hskip -4.8pt Z} \else
       {\hbox {$Z\hskip -4.8pt Z$}}\fi}
\def\qz{\ifmmode {Q\hskip -5.0pt\vrule height6.0pt depth 0pt
       \hskip 6pt} \else {\hbox
       {$Q\hskip -5.0pt\vrule height6.0pt depth 0pt\hskip 6pt$}}\fi}
\def\rz{\ifmmode {I\hskip -3pt R} \else {\hbox {$I\hskip -3pt R$}}\fi}
\def\cz{\ifmmode {C\hskip -4.8pt\vrule height5.8pt\hskip 6.3pt} \else
       {\hbox {$C\hskip -4.8pt\vrule height5.8pt\hskip 6.3pt$}}\fi}
\def\au{{\setbox0=\hbox{\lower1.36775ex%
\hbox{''}\kern-.05em}\dp0=.36775ex\hskip0pt\box0}}
\def\ao{{}\kern-.10em\hbox{``}}

\begin{document}
\begin{titlepage}
\begin{flushright} UWThPh-2000-27\\
ESI-979\\
June 29, 2000
\end{flushright}

\vspace*{2.2cm}
\begin{center}
{\Large \bf Two-dimensional anyons \\[8pt] and the temperature
dependence \\[10pt]of commutator anomalies $^\star$ }\\[30pt]

Nevena Ilieva$^{\ast,\sharp}$\\ [10pt]
{\small\it
Institut f\"ur Theoretische Physik \\ Universit\"at Wien\\
\smallskip
and \\
\smallskip
Erwin Schr\"odinger International Institute\\ for Mathematical
Physics\\}

\vfill \vspace{0.4cm}

\begin{abstract}
The temperature dependence of commutator anomalies is discussed
on the explicit example of particular (anyonic) field operators in
two dimensions. The correlation functions obtained show that
effects of the non-zero temperature might manifest themselves not
only globally but also locally.

\vspace{0.8cm}

\hspace{1.9cm}quantum Hall effect

\end{abstract}
\end{center}

\vfill {\footnotesize

$^\star$ Work supported in part by ``Fonds zur F\"orderung der
wissenschaftlichen Forschung in \"Osterreich" under grant
P11287--PHY;

$^\ast$ On leave from Institute for Nuclear Research and Nuclear
Energy, Bulgarian Academy of Sciences, Boul.Tzarigradsko Chaussee
72, 1784 Sofia, Bulgaria

$^\sharp$ E--mail address: ilieva@ap.univie.ac.at}
\end{titlepage}
\vfill \eject

\setcounter{page}{2}
\section*{Introduction}
Anomalies play an important role in quantum field theory. Originally,
they have been considered at zero temperature and it is therefore of
interest to clarify whether they change upon heating. There is a belief,
strengthened also by particular examples (as the usual Schwinger term in
the current commutator in two dimensions, \cite{HG}), that this is not the
case, that is --- anomalies remain temperature independent also in a thermal
state. A rough argumentation might refer to the fact that anomalies being a
typically local phenomenon, should not be affected by global  features
of the theory, such as the temperature.

That this is not necessarily so will be demonstrated on the explicit example
of particular field operators in two dimensions whose correlation functions
exhibit severe temperature dependence, so that finite-temperature effects
manifest themselves not only globally but also locally.

\section{The anyonic field operators}

In the discussion of Bose--Fermi duality at finite temperature there appear
in a natural way field operators with exotic exchange relations, so anyons
\cite{epj, tmp}
\beq
\Psi_\alpha(x) := \lim_{\ve \ra 0} n_\alpha(\ve) \exp \left[ i
\sqrt{2\pi\alpha} \int_{-\infty}^\infty dy \; \vp_\ve(x-y) j(y)
\right]\, ,
\eeq
with $n_\alpha$ --- some renormalization parameter
and $\vp_\ve (x)$ --- an approximation to the Heaviside function
$$
\lim_{\ve\ra 0}\vp_\ve(x) = \Theta(x), \quad \vp(x) \in H_1\, ,
$$
where $H_1$ is the Sobolev space, $H_1 = \{f: f, f' \in
L^2 \}$. Two special families of such operators are distinguished:
those characterized by statistic parameter $\alpha = 2\cdot2n\pi$
being actually bosons and those characterized by statistic
parameter $\alpha = 2(2n+1)\pi$ being respectively fermions (in
both cases $n \in {\bf Z}$). All operators (1.1) are goverened by
one and the same dynamical equation --- Heisenberg's Urgleichung
(pre-equation) \cite{hei}
 \beq
 \not\!\partial\psi(x) = \lambda\psi(x)\bar\psi(x)\psi(x), \quad
 \lambda = \sqrt{2\pi\alpha} \,.
\eeq
With no bosons present in it at all, Eq.(1.2) represents the ultimate version
of the opinion that fermions should  enter the basic formalism of a fundamental
theory of elementary particles, that is usually taken for granted.

Thus, the coupling constant in Heisenberg's pre-equation turns out
to be related to the statistic parameter, that characterizes
different anyons. Since the latter are uncountably many and live in
orthogonal spaces, the whole Hilbert space becomes nonseparable
and in each of its sectors a different pre-equation holds --- a
feature that certainly cannot be seen by any power expansion in
$\lambda$, which, however, in the above context becomes
problematic itself.

Investigation of  anyonic field operators of the type (1.1) represents by far not only an
academic interest --- such fields might become of importance in solid-state physics, in
problems like quantum wires and FQHE. The relation between the objects there involved
and the field operators (1.1) is rather obvious.

Recall \cite{epj} that being governed by Heisenberg's pre-equation means that
anyons (1.1)  are all solutions to the (finite-temperature) Thirring model
\cite{TM}. On the other hand, it is well known \cite{ML, Hal} that a
one-dimensional  (interacting) electron gas is effectively described by the
Luttinger model --- a Thirring-like model, in which the current-current interaction
is not strictly localized but is characterized by a suitable form-factor \cite{L}
 \beq
 H_{Lut} = i\psi_L^\dag \partial_x\psi_L +
 i\psi_R^\dag\partial_x\psi_R + \int \psi_L^\dag \psi_L(x) V(x-y)
 \psi_R^\dag\psi_R(y) dy \,.
\eeq Actually, for repulsive interactions Hamiltonian (1.3)
is driven by renormalization group flow to the Thirring-model one,
$V(x-y) \ra g \delta(x-y)$, with an effective coupling
constant $g$, so that the
corresponding bosonised Hamiltonian takes the form
\beq H_{Lut} =
\frac{h v_F}{4}\int dx :\left(\frac{1}{v_F^2}(\partial_t\Phi)^2 +
(1+g)(\partial_x\Phi)^2 \right):\,, \eeq where $:\,:$ stands for
Wick's ordering, $v_F$ is the Fermi velocity, $n=\partial_x\Phi$
is the particle dencity and $I=e\partial_t\Phi$ --- the electric
current.

The Hamiltonian (1.4) describes two different physical systems,
a quantum wire and the edges of a Quantum Hall bar. These
systems differ by the form of creation and annihilation
operators of physical electrons. Since the interaction
with the external reservoirs is via electron exchange,
the precise form of the electron operators is crucial for computing
the longitudinal conductance in quantum wires and Hall
conductance in the Quantum Hall setting \cite{AF}.
In the case of quantum wires, one obtains
a one-parameter family of CFT's, with
non-chiral electron operators and non-renormalized (independent on
$g$) conductance value, $\sigma =I/V=e^2/h$. In the
case of QH fluids, the electrons are chiral because the
two edges are separated by a macroscopically big sample.
The Quantum Hall filling fraction is related
to the effective Luttinger coupling by formula,
\beq \nu = \frac{1}{\sqrt{1+g}}\,, \eeq
Hence, the Hall conductance is reduced with
respect to the fundamental unit of $e^2/h$,
$\sigma_H = \nu e^2/h$.
In fact, the incompressibility of the Quantum Hall fluid
requires $\nu^{-1}$ be an odd integer, such that
$\sqrt{1+g} = 2k+1, \,k \in {\bf Z}$. The case of $k=0$
corresponds to the integer Quantum Hall effect, $\nu=1$,
and the noninteracting Luttinger model with $g=0$ is
recovered. In the case of
$k\not=0$ one obtains Laughlin's filling fraction \cite{La} in
the theory of FQHE. In this case, the physical electrons
are identified with Wen's fermions \cite{W, FK} and have an unusual
statistics which depends on the filling fraction.

These edge-excitation operators are special cases of anyons (1.1). However,
in the fermionic case --- $\alpha = 2\pi(2n+1)$, so for Laughlin's states ---
one has to distinguish between fermions, corresponding to $n=0$ and
$n\not=0$. These fields, though locally anticommuting, are severely
different: the former are canonical fields, while the latter are not and this
difference shows up also in their thermal properties as will be argued below.

\section{Commutator anomalies and algebra extensions}

The first known commutator anomaly --- the one in the current commutator
in two dimensions, has been discovered in the thirties by Jordan \cite{J} and
Born \cite{BNN}, in the attempts for construction of neutrino theory of
light. Then the observation has been made that because of the
unboundedness from below of the free-fermion Hamiltonian fermionic
creation and annihilation operators should undergo what is called now a
Bogoliubov transformation. Thus the stability of the system is achieved
but in addition an anomalous term (later called ``Schwinger term")
appeares in the current commutator, so that for the smeared currents
one gets
\beq
[j_f, j_g] = \int_{-\infty}^\infty \frac{dp}{(2\pi)^2} \; p \wt f(p) \wt g(-p) =
\frac{i}{2\pi} \int_{-\infty}^\infty dx f'(x)g(x)
 = i\sigma(f,g)\,,
\eeq
$\sigma(f,g)$ being the symplectic form on the current algebra $\A_c$.
The question is how far is this result state- (so, representation-)
dependent.

The rigorous definition of the anyonic field operators proposed by
us \cite{epj, tmp} allows for a detailed analysis of such questions. It
is based on  the construction of a chain of algebraic inclusions, starting
with the CAR-algebra of bare fermions $\A$,
\beq {\rm CAR}({\it bare})
\subset \pi_\beta(\A)'' \supset\A_c \subset \bar \A_c \subset
\bar\pi_\beta(\bar\A_c)'' \supset {\rm CAR}({\it dressed})\,.
\eeq
The shift $\tau_t$ is an automorphism of $\A$ which has KMS-states
$\omega_\beta$ and associated representations $\pi_\beta$. In
$\pi_\beta(\A)''$ one finds bosonic modes $\A_c$ with an algebraic
structure independent on $\beta$. The crucial ingredient needed
so far was the appropriately chosen state. We make use of the
KMS-state (which is unique for the shift over the CAR algebra). Another
possibility would be to introduce the Dirac vacuum (filling all
negative energy levels in the Dirac sea). This is  what has been
originally done in the thirties \cite{J, BNN}, and recovered later
by Mattis and Lieb \cite{ML} in the context of the Luttinger
model. However, an important detail might be overseen that way:
symplectic structure (2.1) though formally independent on $\beta$
(see also \cite{HG}),  for $\beta < 0$ changes its sign, $\sigma
\to -\sigma$, and for $\beta = 0$ (the tracial state) becomes
zero, that is
\beq
[j_f,j_g] = i \sigma(f,g)  
 = \left\{ \ba{cl} \int f'g \,\, & \quad \mbox{ for } \beta>0\\ 0 &
 \quad\mbox{ for } \beta=0\\ -\int f'g & \quad\mbox{ for }
 \beta<0 \ea \right.
\eeq

\noindent
{\it Remark.\/} It is the parity $P$ (which suffers a destruction on the
passage from the CAR-algebra to the current algebra \cite{dub,
tmp}) that relates the states corresponding to positive and
negative temperatures
$$
\omega_{-\beta} = \omega_\beta \circ P\,.
$$

\vspace{12pt}

In the construction of algebraic chain (2.2) we make use of the
$\tau$-KMS states, which are translation-invariant equilibrium states
at an inverse temperature $\beta$. On $\A_c$ such a state is given by
the two-point function
$$
\omega(j(f)j(g)) = \int dxdy \,w(x-y)f(x)g(y)\, ,
$$
with a kernel
\beq
w(x-y) =-\lim_{\ve\ra0^+}\,\frac{1}{(2\pi)^2\,\sh^2(x-y-i\ve)}\, .
\eeq

Recall that for Weyl operators, the multiplication law replaces relation (2.1)
\beq
e^{ij(f)} \; e^{ij(g)} =
e^{\frac{i}{2} \sigma(g,f)} \; e^{ij(f+g)}\,.
\eeq
with $\sigma$ --- the symplectic form on the algebra.
The expectation of the Weyl operators is given by
\beq
\omega(e^{ij(f)}) = e^{-{1\over 2}\langle f | f\rangle}\, ,
\eeq
where the scalar product $\langle f | g\rangle$ defines the one-particle
real Hilbert space $h$ of the $f$'s. For consistency, it has to satisfy
$$
\sigma(f|g) = \left(\langle g|f\rangle - \langle f|g\rangle\right)/2\, .
$$
Eqs.(2.5),(2.6) imply
$$
\omega(e^{ij(f)}e^{ij(g)}) = \exp{\left\{-{1\over 2}
\left[ \langle f | f \rangle+\langle g | g \rangle +
2\langle f | g \rangle \right]\right\}}\, ,
$$
or generally
\beq
\omega(\prod_k e^{ij(f_k)}) = \exp{\left\{-{1\over 2}
\left[\dsum_k \langle f_k | f_k \rangle +
2\dsum_{k<m} \langle f_k | f_m \rangle \right]\right\}}\, .
\eeq

\section{The KMS correlation functions}

The anyons (1.1) are Weyl operators for which
the smearing function is $f^x_\alpha(y) = \sqrt{2\pi\alpha}
\Theta(x-y)$.  So far $\A_c$ was defined for $j_f$'s with $\, f \in C_0^\infty$,
for instance, that is with functions which vanish for $x\ra\pm\infty$.
The structure of $\A_c$ is determined by the
symplectic form $\sigma(f,g)$ (2.1) which is actually well
defined also for the Sobolev space, $\sigma(f,g) \ra \sigma(\bar
f, \bar g), \, \bar f, \bar g \in H_1, \, H_1 = \{f : f,f' \in
L^2\}\,$. Also $\bar \omega_\beta$ can be extended to $H_1$, since
$\,\bar \omega_\beta(e^{ij_{\bar f}}) > 0\,$ for $\,\bar f \in
H_1$.  Thus, the symplectic form (2.1) may be given a sense
for functions that tend to a constant, however they cannot be reached
as limits of functions from $\C_0^\infty$. For instance,
\beq
\Phi_{\delta,\ve}(x)  := \vp_\ve(x)
- \vp_\ve(x + \delta) \in H_1,
\eeq
$$
 \lim_{\delta\ra\infty \atop \ve\ra 0}\Phi_{\delta,\ve}(x)=\Theta(x)
$$
with
\beq
\vp_\ve(x) :=
\left\{\ba{cl} 1 & \mbox{ for } x \leq -\ve \\ -x/\ve & \mbox{ for }
- \ve \leq x \leq 0 \\ 0 & \mbox{ for } x \geq 0 \ea \right.\,.
\eeq
does not work, since $\sigma(\Phi_{x,\delta}, \Phi_{x',\delta'})$ depends on the
order in which the limits $\delta, \delta' \ra\infty$ are taken and only for $\delta=
\delta'\ra\infty$ we get the desired result $i\,\sgn(x-x')$. Since this appears
in the $c$-number part, in no representation can $j(\Phi_{x,\delta})$ converge
strongly. Nevertheless, for functions with the same (nontrivial) asymptotics
at, say, $x\ra\infty$ and whose difference $\in h$ one can succeed  in getting
the expectation values as limits. In the case when different anyonic
contributions ``compensate" each other a nontrivial result is  obtained due
to the cancellation of the infrared diveregencies.
In terms of $\Psi$'s this means that the expectation value of a product
of $\Psi$'s and $\Psi^*$'s is different from zero only if there are as many
$\Psi$'s as $\Psi^*$'s, equivalently --- if the ``total" statistic parameter of
creation operators equals the one of annihilation operators, for instance
$\langle \Psi^*_1\Psi^*_1\Psi_4\rangle\not=0$, or otherwise that they lead to
orthogonal sectors of the enlarged Hilbert space.

Next we have to cope with the ultraviolet problem. Here the situation is different
since we do not need a larger Hilbert space to this end. The solution is already
contained in the usage of a smeared step function in (3.1). A trivial integration
then gives, e.g. for the $\alpha$ two-point function $(F_{x,\ve}^\alpha :=
\sqrt\alpha  \vp_\ve(x))$
\beqa
\omega(\Psi^*_\alpha(x)\Psi_\alpha(y)) = \omega(e^{-ij(F_{x,\ve}^\alpha)}
e^{ij(F_{y,\ve}^\alpha})\no \\[6pt]
= \left( \frac{i\ve}{\sh{\frac{\pi(x-x'-i\ve)}{\beta}}}\right)^{(\alpha/2\pi)} \, .
\eeqa

The divergence for $\ve\ra 0$ remains and determines the neccessary
renormalization of the operators $\Psi_\alpha$ \cite{tmp}.
Thus, for $\alpha = 2\pi$ the renormalization parameter should be
$n_{2\pi}(\ve) = (2\pi\ve)^{-1/2}$, so that the fermion KMS two-point
function is recovered,
and for the general case of an arbitrary $\alpha\,$: $\,n_\alpha(\ve) =
(2\pi\ve)^{-\alpha/4\pi} = n_{2\pi}^\alpha(\ve)$.

For all $\alpha$'s the two-point function (for $x > x'$ and $\beta =
\pi$)
\beq
\langle \Psi_\alpha^*(x) \Psi_\alpha(x') \rangle_\beta =
\langle \Psi_\alpha(x) \Psi_\alpha^*(x') \rangle_\beta =
\left(\frac{i}{2\beta \sinh(x-x')}\right)^{\alpha/2\pi} =: S_\alpha(x-x')
\eeq
has the desired properties
\begin{enumerate}
\item \Un{Hermiticity}:
$$
S_\alpha^*(x) = S_\alpha(-x) \, \Longleftrightarrow \, \langle
\Psi_\alpha^*(x)\Psi_\alpha(x')\rangle_\beta^* = \langle
\Psi_\alpha^*(x')\Psi_\alpha(x)\rangle_\beta\,;
$$
\item \Un{$\alpha$-commutativity}:
$$
S_\alpha(-x) = e^{i\alpha/2}S_\alpha(x) \, \Longleftrightarrow \,
\langle\Psi_\alpha(x')\Psi_\alpha^*(x)\rangle_\beta =
e^{i\alpha/2}\langle\Psi_\alpha^*(x)\Psi_\alpha(x')\rangle_\beta \, ;
$$
\item \Un{KMS-property}:
$$
S_\alpha(x) = S_\alpha(-x+i\pi) \, \Longleftrightarrow \,
\langle\Psi_\alpha^*(x)\Psi_\alpha(x')\rangle_\beta =
\langle\Psi_\alpha(x')\Psi_\alpha^*(x+i\pi)\rangle_\beta\, .
$$
\end{enumerate}

For $\alpha = 2\pi$ Eq.(5.7) reads
\beq
\langle\Psi_{2\pi}^*(x)\Psi_{2\pi}(x')\rangle_\beta = \lim_{\ve\ra
0_+}\frac{i}{2\beta\,\sinh \frac{\pi(x-x'-i\ve)}{\beta}}\, .
\eeq

For $f(y)\ra\Theta(x-y)$, $\,g(y')\ra -\Theta(x'-y')$ nothing changes so
we
verify the relation (5.2) $\,\leftrightarrow\,$ (5.3),
$$
\langle\Psi_{2\pi}^*(x)\Psi_{2\pi}(x')\rangle_\beta =
\langle\Psi_{2\pi}(x)\Psi_{2\pi}^*(x')\rangle_\beta\, .
$$

For $\alpha = 4\pi$ we get like for the $j$'s
\beq
\langle\Psi_{4\pi}^*(x)\Psi_{4\pi}(x')\rangle_\beta =
-\frac{1}{\left(2\beta\,\sinh{\frac{\pi(x-x'-i\ve)}{\beta}}\right)^2}\, ,
\eeq
whereas for $\alpha = 6\pi$ we get a different kind of fermions
\beq
\langle\Psi_{6\pi}^*(x)\Psi_{6\pi}(x')\rangle_\beta =
-\frac{i}{\left(2\beta\,\sinh{\frac{\pi(x-x'-i\ve)}{\beta}}\right)^3}\, .
\eeq

This shows that local anticommutativity alone does not guarantee the
uniqueness of the KMS-state, one needs in addition the CAR-relations. The
$\Psi_\alpha$'s, $\,\alpha \in 2(2{\bf N}+1)\pi\,,\,$ describe an infinity
of inequivalent fermions.

Thus we find a kind of a quantum selection rule for our original
dynamical equation --- Heisenberg's pre-equation. It admits
fermionic solutions only for $\lambda = \sqrt{2\pi}$. Of course
one could for any $\lambda$ enforce Fermi statistics by
renormalizing the bare fermion field $\psi \ra \sqrt{Z}\;\psi$, $j
\ra Zj$ with a suitable $Z(\lambda)$ but this just means pushing
factors around. Also for the series of values $\lambda =
\sqrt{2(2n+1)\pi}$, $n\in {\bf Z^+}, \, n\not= 0$ one gets
anticommuting fields as solutions, however they are noncanonical.

These non-canonical Fermi-fields are similar to Wen's fermions
$$
\langle \psi(z)\psi^\dag(w)\rangle \sim
\frac{1}{(z-w)^{2n+1}}
$$
which correspond to Laughlin's plateaux in the theory of the FQHE, but
considered at finite temperature. For a detailed analysis of this relation we
refer to \cite{INT2}.

In complete analogy, for the $n$-point function we obtain
\beqa
&&\omega\left(\Psi_\alpha^*(x_1)\dots\Psi_\alpha^*(x_n)
\Psi_\alpha(y_n)\dots\Psi_\alpha(y_1)\right) \no\\[8pt]
& = &
\frac{\dprod_{k>l}(\sh(x_k-x_l-i\ve))^{\alpha/2\pi}
\dprod_{k>l}(\sh(y_k-y_l-i\ve))^{\alpha/2\pi}}
{\dprod_{k,l} \left(-2\pi i \,\sh(x_k-y_l-i\ve)\right)^{\alpha/2\pi}}\,.
\eeqa
The exact exchange relations are hidden in the factor $i^\alpha$ in (3.7) and
we shall return to their detailed analysis later on.

For the case $\alpha = 2\pi$  with the help of Cauchy's determinant formula
it can be rewritten as
\beq
 \frac{\dprod_{i>k}
\sh(x_i-x_k-i\ve)\dprod_{i>k}\sh(y_i-y_k-i\ve)}
{\dprod_{i,k}\sh(x_i-y_k-i\ve)}= \Det\frac{1}{\sh(x_i-y_k-i\ve)} \,,
\eeq
so we recover the original free fermions with all their characteristic
properties.

\section{Temperature dependence of the $\alpha$-commutator}

For the thermal expectation of the $\alpha$-commutator one finds
\beqa
\omega_\beta([\Psi^*(x), \Psi(x')]_\alpha) &=&
-i\left(-\frac{1}{2\beta\,\sinh{\frac{\pi(x-x'-i\ve)}{\beta}}}\right)
^{\alpha/2\pi}
\no \\
&+&
i\left(-\frac{1}{2\beta\,\sinh{\frac{\pi(x-x'+i\ve)}{\beta}}}\right)
^{\alpha/2\pi}
\eeqa
and for $\ve \ra 0$ a distribution is obtained where only the leading
singularity is temperature independant, namely
$$
\lim_{\ve\ra 0}
\left\{\left(\frac{1}{x-x'-i\ve}\right)^{\alpha/2\pi}
- \left(\frac{1}{x-x'+i\ve}\right)^{\alpha/2\pi}\right\} =
\, ?
$$
For example, for $\alpha = 6\pi$ this expectation is
$$
\langle -[\Psi_{6\pi}^*(x),\Psi_{6\pi}(x')]_\alpha\rangle_\beta
= \langle \, [\Psi_{6\pi}^*(x),\Psi_{6\pi}(x')]_+\rangle_\beta =
-\frac{1}{8\pi^2}\left(\delta''(x-x') -
\frac{\pi^2}{\beta^2}\delta(x-x')\right)\,.
$$

The commutation relations themselves exhibit an operator structure
\cite{INT}. Thus, while for $\alpha = 2\pi$ the canonical commutation
relations are recovered, for $\alpha = 4\pi$ the commutator reads
\beq
[\Psi_{4\pi}^*(x), \Psi_{4\pi}(y)] ={i\over 2\pi} \delta'(x-y) +i\sqrt 2\delta(x-y)j(x)\,,
\eeq
which is essentially the Frenkel--Kac representation of the $su(2)$
current algebra, and for $\alpha = 6\pi$ the anticommutator takes the
form
\beqa
[\Psi_{6\pi}^*(x), \Psi_{6\pi}(y)]_+ &=&-{1\over 8\pi^2}[\delta''(x-y) -2\pi\sqrt 3
\delta'(x-y)j(x) + 12 \pi^2\delta(x-y)j^2(x)] \no \\[4pt]
&+& {1\over 8\beta^2}\delta(x-y)\,.
\eeqa
This demonstrates once again the non-canonical nature
already of the first extension classes --- bosonic, as well as fermionic
--- beyond the initial Fermi-algebra. The temperature dependece
appears with the first non-canonical Fermi-class. Since
the algebra of the $6\pi$-fermions $\A_{(3)} := \{\Psi_\alpha:
\alpha = 6\pi\}$ is a subalgebra of  $\A_{(1)} := \{\Psi_\alpha:
\alpha = 2\pi\}$, the algebra of the original fermions, this shows
that representations that correspond to different temperatures
are inequivalent not only globally, but also locally. In turn, this
means that the temperature effects in this construction become
locally observable.

\section{Conclusions and outlook}
We have discussed the temperature dependence of correlation
functions and exchange relations of the two-dimensional anyons
constructed in \cite{epj, tmp}. For the (infinity of inequivalent)
anticommuting fields  that are present among them, this dependence
means a loss of the local normality of representations corresponding
to different temperatures, hence --- (already local) observability of
the temperature effects. Consideration of these noncanonical
Fermi-fields in detail is of particular interest since their relation to
Wen's fermions allows for a rigorous analysis of the possibility for a
second-quantization picture of the  fractional quantum Hall effect
\cite{sq}.

\section*{Acknowledgements}

I appreciate the suggestive discussions with A. Alekseev and
L. Georgiev and the useful remarks of K.-H. Rehren on the matter of
this note and the fruitful collaboration with H. Narnhofer and
W. Thirring on various related topics.

The hospitality and financial  support of the International Erwin
Schr\"odinger Institute for Mathematical Physics where part of the
research has been performed, are gratefully acknowledged. This work
has been supported in part also by ``Fonds zur F\"orderung der
wissenschaftlichen Forschung in \"Osterreich" under grant
P11287--PHY.

\end{document}